\documentclass[a4paper,fleqn,usenatbib]{mnras}

\usepackage{newtxtext,newtxmath}
\usepackage[T1]{fontenc}
\usepackage{ae,aecompl}
\usepackage{graphicx}	
\usepackage{amsmath}	
\usepackage{amssymb}	
\usepackage{natbib}
\usepackage{xcolor}

\title[A Toy Model for the Dynamical Discrepancies on Galactic Scales]{A Toy Model for the Dynamical Discrepancies on Galactic Scales}

\author[J. Petersen and M. Rosenlyst]{
Jonas Petersen$^{1}$\thanks{E-mail: petersen@cp3.sdu.dk}
and Martin Rosenlyst$^{1}$
\\
$^{1}$ Centre for Cosmology and Particle Physics Phenomenology, University of Southern Denmark, Campusvej 55, DK-5230 Odense M, Denmark}

\date{Accepted 2019 September 12. Received 2019 September 12; in original form 2019 June 25}

\pubyear{2019}

\begin{document}
\maketitle

\begin{abstract}
\noindent In this study a simple toy model solution to the missing gravity problem on galactic scales is reverse engineered from galactic data via imposing broad assumptions. It is shown that the toy model solution can be written in terms of baryonic quantities, is highly similar to pseudo-isothermal dark matter on galactic scales and can accommodate the same observations. In this way, the toy model solution is similar to MOND modified gravity in the Bekenstein-Milgrom formulation. However, it differs in the similarity to pseudo-isothermal dark matter and in the functional form. In loose terms, it is shown that pseudo-isothermal dark matter can be written in terms of baryonic quantities. The required form suggests that it may be worth looking into a mechanism that can increase the magnitude of the post-Newtonian correction from general relativity for low accelerations.
\end{abstract}


Since the 1930s astronomical observations have made it increasingly clear that the dynamical behaviour of the universe is poorly understood. These observations include the virialization of the Coma cluster \citep{Zwicky:1933gu}, rotation curves \citep{Bosma,Rubin:1980zd}, lensing of galaxy clusters \citep{Mellier}, observation of cluster mergers \citep{Clowe:2006eq} and large scale structure surveys \citep{Dodelson:2006zt}. The discrepancy between the observed and predicted dynamics can be remedied by introducing a significant, as of yet unobserved mass component (dark matter) or modifying the theory responsible for the predicted dynamics - with the former being the vastly more explored option.

Analyses of rotation curve data have historically been the source of much debate (e.g. \citep{deBlok:1998bh,Sanders:2002ue,Sanders:2007rg,McGaugh:2016leg,Petersen:2017klw,Frandsen:2018ftj,Rodrigues:2018duc,Li:2018tdo,Petersen:2019obe,chang2019,Kroupa:2018kgv,Rodrigues:2018lvw,2018NatAs...2..924M,Sanders:2018mnk,Tian2019}) since \citep{Tully:1977fu} inferred a close relation between the visible baryonic matter and behaviour of galaxies at large radii - supposedly dominated by dark matter. The debate spawned the idea of modified classical dynamics, introduced by \citet{Milgrom:1983ca}, as an alternative to the dark matter hypothesis. Milgrom coined the non-relativistic modification of gravitational dynamics, relevant on galactic scales, modified Newtonian dynamics (MOND). He defined two classes, namely MOND modified inertia and MOND modified gravity \citep{Milgrom:1983ca}. The former model class was reinvigorated after \citet{McGaugh:2016leg,Lelli:2017vgz} found, in their analysis of rotation curve data from the SPARC database \citep{Lelli:2016zqa}, that data from it globally follow an analytical relation (dubbed the RAR) that carries the features of MOND modified inertia. The results of \citet{McGaugh:2016leg,Lelli:2017vgz} even inspired dark matter model builders to reproduce the RAR, with the unique characteristics of MOND modified inertia, within the dark matter hypothesis e.g. ~\citet{Chashchina:2016wle,Edmonds:2017zhg,Dai:2017unr,Cai:2017buj,Berezhiani:2017tth}. Subsequently, the discussion surrounding MOND modified inertia has continued, with \citep{Li:2018tdo} coming out with evidence in support and \citep{Petersen:2017klw,Frandsen:2018ftj,Petersen:2019obe} the contrary.\newline\newline 
\noindent As the solution to the missing gravity problem continues to elude physicists, good ideas continue to grow in importance and new inspirations become of increasingly greater value. In this article, a toy model is derived via reverse engineering from galactic rotation curve data. The purpose of the toy model is to inspire dark matter and modified gravity model builders by highlighting a curious connection between the toy model and the baryonic mass components on galactic scales and analysing what this could mean.\bigskip

\section{Formulating the Missing Gravity Problem for Rotation Curves}
In relation to rotation curves, the missing gravity problem can be neatly formulated analytically via Newtonian mechanics as follows: Solving the Poisson equation outside a given mass distribution (i.e. solving the Laplace equation; $\vec{\nabla}^2\Phi_k=0$) in cylindrical coordinates yields the potentials
\begin{equation}
	\Phi_k=c(k)e^{-k|z|}J_0(kr),
	\label{pote}
\end{equation}
where $J_0(kr)$ is the Bessel function of the first kind of zeroth order and the coordinate system is arranged such that the $z$-axis is perpendicular to the galactic plane. These solutions are valid outside a given mass distribution and are as such not mass specific. The mass specific solution is developed with $\Phi_k$ as basis functions. This leads to the total (abbreviated "tot") centripetal acceleration 
\begin{equation}
	\vec{g}_{tot}=-\int dk S(k)\vec{\nabla}\Phi_k,
\end{equation}
where $S(k)$ is a weight function to be determined by the mass distribution. Assuming the mass is distributed in a plane~\citep{binney},
\begin{eqnarray}
	\Sigma(r)&=&-\frac{1}{2\pi G_N}\int dk S(k)c(k)kJ_0(kr).
	\label{bin}
\end{eqnarray}
The procedure in Newtonian mechanics is then to assume that $\Sigma(r)$ is the mass distribution of the baryonic matter only. By doing so $S(k)c(k)$, can be determined via Hankel transformation. The existence of dark matter or some modification of Newtonian gravity can be inferred when it is concluded that the Newtonian model does not agree well with data at large radii. In particular, $v_{tot}(r)\sim \text{const}$ is required by observations \citep{Bosma,Rubin:1980zd} while $v_{N}(r)\sim r^{-\frac{1}{2}}$ at large radii from Newtonian mechanics. Hence, it is clear that, in this formalism, $\Sigma(r)$ cannot only be the visible (baryonic) mass. $\Sigma(r)$ should be a sum of the baryonic mass as well as some additional mass term, $\Sigma_{m}$, where "m" signifies "missing" mass. This missing mass term can be interpreted as either arising from dark matter or from \emph{some} modification of Newtonian gravity. The total surface density can then be written as
\begin{equation}
	\Sigma_{tot}(r)=\Sigma_{N}(r)+\Sigma_{m}(r) \Rightarrow \vec{g}_{tot}=\vec{g}_{N}+\vec{g}_m.
	\label{sig}
\end{equation}

\section{The Toy Model}
The toy model consists of an analytical expression for the radial component (Notation: $g_{\dots,r}=|(1\,0\,0)\cdot\vec{g}_{\dots}|=\frac{v_{\dots}^2}{r}$) of $\vec{g}_m$ derived via reverse engineering from galactic rotation curve data. The reverse engineering process itself consists of three broad assumptions motivated by the data;
\begin{enumerate}
	\item $v_{tot}(r)\sim \text{const}$ at large radii \citep{Bosma,Rubin:1980zd}.
	\item $g_{m,r}$ should not diverge at $r\rightarrow 0$.
	\item $v_{tot}\sim v_{N}$ at small radii~\citep{Petersen:2019obe}, corresponding also to the maximum disk approximation \citep*[e.g.][]{Albada1985}.
\end{enumerate}
All three assumptions are rather uncontroversial, albeit the third one less so. The third assumption represents the maximum disk approximation or equivalently the cored dark matter profiles. The slight controversy of this assumption does not persist in there existing galaxies which uphold this assumption, but in the assumption that all galaxies do. \citet{Petersen:2019obe} recently motivated this assumption in an analysis of the gas dominated galaxies of the SPARC database and indicated that this might in fact be the case in all galaxies.\newline

\noindent Assumption 1 can be accommodated by requiring $g_{m,r}\sim r^{-1}$ at large radii. In order to also uphold assumption 2, a first ansatz could be
\begin{equation}
	g_{m,r}|_{z=0}\sim \frac{1-e^{-\frac{r}{r_{m}}}}{r},
	\label{eq1}
\end{equation}
where $r_{m}$ is the scale length of the missing mass and $g_{m,r}$ is evaluated at $z=0$ since this denotes the galactic plane to which the rotation curve data refer. Equation \eqref{eq1} has $\lim\limits_{r\rightarrow 0}(g_{m,r})\neq0$, which violates assumption 3. A re-evaluated ansatz could therefore be
\begin{equation}
	g_{m,r}|_{z=0}= \Phi_m^0\bigg[\frac{1-e^{-\frac{r}{r_{m}}}}{r}-\frac{e^{-\frac{r}{r_{m}}}}{r_{m}}\bigg],
	\label{G}
\end{equation}
where $\Phi_m^0$ is a constant of proportionality that is in general a function of scale length and mass. Equation \eqref{G} is not unique in that there exist other parametrizations that uphold the three assumptions. What makes this particular parametrization interesting is its relation to the baryonic matter. To uncover this relation, assume that $g_{m,r}$ exists in the same function space as $g_{N,r}$. This is equivalent to taking
\begin{equation}
	\begin{split}
		&S(k)c(k)=S_{N}(k)+S_{m}(k)\\
		&\,\,\Rightarrow g_{m,r}=-\int_{0}^{\infty}dk S_{m}(k)kJ_1(kr)e^{-k|z|},
	\end{split}
	\label{Hankel}
\end{equation}
where $S_{N}$ is the weight function obtained from the baryonic mass distribution using Newtonian dynamics and $J_1(kr)$ is the Bessel function of first kind of first order. Taking $z=0$ in equation \eqref{Hankel}, using equation \eqref{G} and  Hankel transforming reveals 
\begin{eqnarray}
	S_{m}(k)&=& -\Phi_m^0\int_{0}^{\infty}dr'r'J_1(kr')\bigg[\frac{1-e^{-\frac{r'}{r_{m}}}}{r'}-\frac{e^{-\frac{r'}{r_{m}}}}{r_{m}}\bigg]\nonumber\\
	&=&\frac{-1}{(1+(kr_{m})^2)^\frac{3}{2}}\frac{\Phi_m^0}{k}.
	\label{g1}
\end{eqnarray}
Compare this to the weight function for an exponential baryonic disk \citep{binney}:
\begin{equation}
	S_{d}(k)=-\frac{G_Nm_d}{(1+(kr_d)^2)^{\frac{3}{2}}},
	\label{g2}
\end{equation}
where $r_d$ and $m_d$ are the scale length and mass of the baryonic disk, respectively. The difference in powers of $k$ between equations \eqref{g1} and \eqref{g2} can be captured by an indefinite integral over $|z|$. 
\begin{equation}
	g_{tot,r}= g_{N,r}-\frac{\Phi_m^0}{G_Nm_d}\int d|z|\partial_r\Phi_{d}(r,z,r_m).
	\label{pot}
\end{equation}
In order for assumption 1 to appear from data, both the magnitude of $g_{m,r}$ and the involved scale lengths must -- to some degree -- be fine tuned towards the baryonic ones. Hence, as an ansatz it is reasonable to let $r_m\rightarrow r_d$ such that
\begin{equation}
	g_{tot,r}=g_{N,r}-\frac{1}{r_c}\int d|z| \partial_r\Phi_{d},\\
	\label{pot1}
\end{equation}
where it is now understood that $\Phi_d$ is the baryonic disk potential with the baryonic scale length and magnitude and
\begin{equation}
	\frac{1}{r_{c}}\equiv\frac{\Phi_m^0}{G_Nm_d}.
\end{equation} 
By modelling the baryonic disk and gas as exponential disks \citep{binney}, a sum can be introduced into Equation \eqref{pot1} in a straightforward manner. In cases where there is no baryonic bulge, then $\sum_i\int d|z| \partial_r\Phi_{d}^i=\int d|z| g_{N,r}$. A priori there is no reason why the missing mass term should not depend on the baryonic bulge. However, introducing a baryonic bulge is consistent with assumptions 1-3. The only possible impact of a bulge is in identifying Equation \eqref{G} with baryonic quantities. Since Equation \eqref{G} is defined for $z=0$, a bulge component can be consistently added with the requirement that it vanishes in the limit $z\rightarrow 0$. This can be handled by the integration constant of the indefinite integral, allowing Equation \eqref{pot1} to be written 
\begin{equation}
	g_{tot,r}=g_{N,r}-\frac{1}{r_c}\int d|z| g_{N,r},
	\label{potx}
\end{equation}
where 
\begin{equation}
	g_{N,r}=\sum_i\partial_r\Phi_{d}^i+\partial_r\Phi_{b}
\end{equation}
with $\lim\limits_{|z|\rightarrow \infty}(\int d|z| g_{N,r})=0$ and $\lim\limits_{|z|\rightarrow 0}(\int d|z| \partial_r\Phi_{b})=0$ imposed.

\section{Comparing the Toy Model to Rotation Curve Data}
In comparing the toy model to data, the geometry in $(g_{N,r},g_{tot,r})$-space ($g2$) will be investigated. $g2$-space -- used by \citet{Milgrom:1983ca} to write the now famous mass discrepancy acceleration relation (MDAR) -- presents several advantages compared to more conventional spaces (e.g. the classical $v(r)$ plane) in that it highlights the details of the solution to the missing gravity problem. For example, the difference in the $v(r)$ plane for a NFW dark matter density profile and a pseudo-isothermal dark matter density profile, both fitted to data, can be difficult to see. In $g2$-space however, the difference is very explicit and clear (Figure \ref{fig:3}).\newline\newline

\begin{figure}
	\centering
	\includegraphics[width=1\columnwidth]{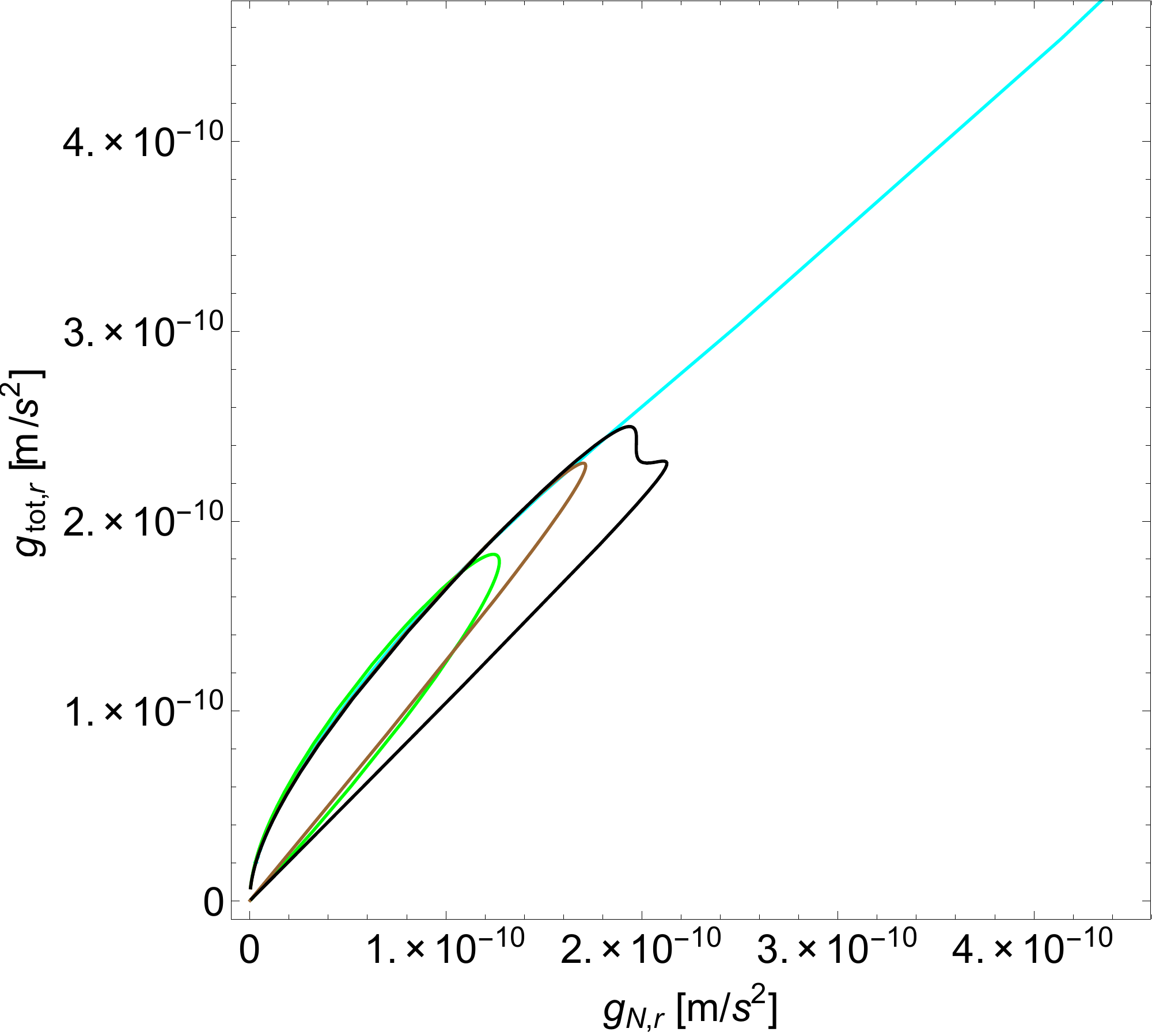}
	\caption{The $g2$-space geometries obtained from using Equations \eqref{18} and \eqref{19} for the baryonic content. The different curves denote different baryonic content; green is an exponential disk, cyan is an exponential disk and a de Vaucouleurs bulge, brown is a Miyamoto-Nagai disk and black is a Miyamoto-Nagai disk and a Plummer bulge. For the figure $z=0$, $r\in [0.01\,\text{kpc},100\,\text{kpc}]$, $r_b=\tilde{r}_b=0.5\,\text{kpc}$, $m_d=5.72\times 10^{40}\,\text{kg}$, $m_b=1.43\times 10^{39}\,\text{kg}$, $a_d=r_d=3\,\text{kpc}$, $b_d=0.1\,\text{kpc}$, $\tilde{\Sigma}_0=2.4\,\frac{\text{kg}}{\text{m}^2}$, $\Sigma_0=1.06\,\frac{\text{kg}}{\text{m}^2}$ and $r_c=\sqrt{\frac{f G_N(m_d+m_b)}{cH_0}}$ with $f$ a free parameter here taken to be $2\pi$ and $H_0$ being the Hubble rate.}
	\label{fig:1}
\end{figure}

\noindent There exist various different models of the baryonic matter in rotationally supported galaxies. The most common are \citep{Sofue:2013hja}:
\begin{enumerate}
	\item Exponential disk and de Vaucouleurs bulge with surface mass densities
	\begin{equation}
		\begin{split}
			\Sigma_d(r)&=\Sigma_0e^{-\frac{r}{r_d}},\\
			\Sigma_b(r)&=\tilde{\Sigma}_0e^{-\kappa\big((\frac{r}{r_b})^{1/4}-1\big)},
		\end{split}
		\label{18}
	\end{equation}
	where $\kappa=7.6695$ \citep{binney}.
	\item Miyamoto-Nagai disk and Plummer bulge with potentials
	\begin{equation}
		\begin{split}
			\Phi_d(r,z)&=-\frac{G_Nm_d}{\sqrt{r^2+(a_d+z)^2}},\\ 
			\Phi_b(r,z)&=-\frac{G_Nm_b}{\sqrt{r^2+z^2+\tilde{r}_b^2}},
		\end{split}
		\label{19}
	\end{equation}
	where both potentials are in cylindrical coordinates and $a_d,\tilde{r}_b$ are scale lengths.
\end{enumerate}
In relation to the geometry in $g2$-space, it is interesting to note the differences introduced by the bulge profiles. Even though the centripetal acceleration of the de Vaucouleurs bulge, $g^{V}_{bul}$, vanishes for both $r\rightarrow0$ and $r\rightarrow\infty$, the maximum of $g^{V}_{bul}$ is located at incredibly small radius (few $pc$). This radius is well below conventionally sampled radii~\citep{Lelli:2016zqa} and so for the sampled radii a de Vaucouleurs-like bulge will cause $g_{N,r}$ to increase toward small radii and thus force the $g2$-space geometry to be single valued (see Figure \ref{fig:1}). 
Similarly to the de Vaucouleurs bulge, the centripetal acceleration from the Plummer bulge, $g^{P}_{bul}$, vanishes for both $r\rightarrow0$ and $r\rightarrow\infty$. However, contrary to the de Vaucouleurs bulge, the maximum of $g^{P}_{bul}$ is located at radii often sampled by data (few $kpc$). This means the $g2$-space in this case is double valued in the range of sampled radii (see Figure \ref{fig:1}). Without bulges, the exponential disk and the Miyamoto-Nagai disk gives rise to similar, $\sim$elliptical geometries in $g2$-space -- with the latter a bit more "squeezed" than the former (see Figure \ref{fig:1}).\newline
In relation to the $g2$-space geometry seen in the SPARC database \citep{Lelli:2016zqa,Frandsen:2018ftj,Petersen:2019obe}, it is interesting to note that the majority of galaxies\footnote{Here the galaxy selection criteria of \citet{Frandsen:2018ftj,Petersen:2019obe} are applied, bringing the number of galaxies down to $152$ from the $175$ listed in the database.} ($121/152$) have no bulge. Hence, the overall $g2$-space geometry is expected to be dominated by the disk geometries shown in figure \ref{fig:1} - this is indeed what is found in \citep{Petersen:2017klw,Petersen:2019obe} and, to a lesser extent, in \citep{Frandsen:2018ftj}. In \citep{Frandsen:2018ftj,Petersen:2019obe} it is shown that galaxies can in general be grouped according to whether data for the $g2$-space geometry curves leftward ($r_{tot}<r_{N}$), rightward ($r_{tot}>r_{N}$) or nowhere ($r_{N}=r_{tot}$) for decreasing radius, where $r_{N}$ and $r_{tot}$ denote the radii corresponding to the maxima in $g_{N}$ and $g_{tot}$, respectively. \citet{Frandsen:2018ftj} show that larger subsets curve rightward ($73/152$ galaxies) and nowhere ($46/152$ galaxies) whereas a smaller subset curves leftward ($33/152$ galaxies), leading to an overall geometry $\sim$ nowhere with a slight rightward inclination. \citet{Petersen:2019obe} show that gas dominated galaxies -- which to leading order have $g2$-space geometries independent of the mass to light ratios -- overall display a pronounced rightward geometry, indicating that leftward (and possibly nowhere) galaxies could be an artefact of an underlying radial dependency of the mass to light ratios (e.g. as suggested by \citep{Kroupa:2018kgv}) not accounted for in \citet{Frandsen:2018ftj}. Equation \eqref{potx} relies on this assumption as it cannot accommodate leftward or nowhere geometries for which data are sampled at $r<r_{N}$ (see Figure \ref{fig:1}). Via de Vaucouleurs-like bulges, Equation \eqref{potx} can accommodate nowhere geometries which are not sampled at $r<r_{N}$. The toy model can be adjusted to accommodate all types of geometries by introducing more complexity, but this makes the connection with the baryonic matter less pronounced. Since the need for additional complexity is still under debate, further considerations in this direction are beyond the scope of our toy model.\newline

\section{Comparing the Toy Model to Dark Matter Models}
\begin{figure}
	\centering
	\includegraphics[width=0.65\columnwidth]{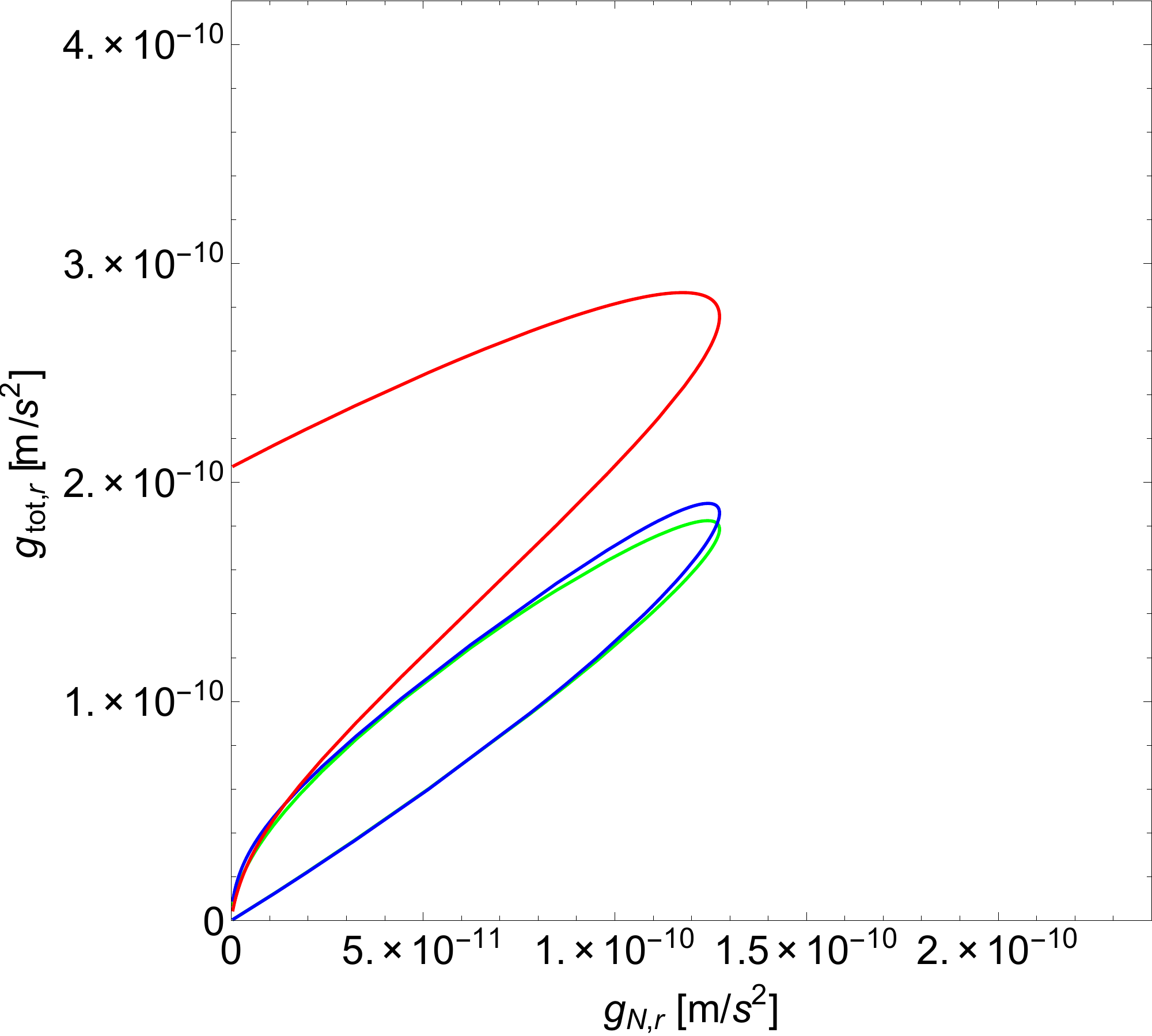}
	\includegraphics[width=0.65\columnwidth]{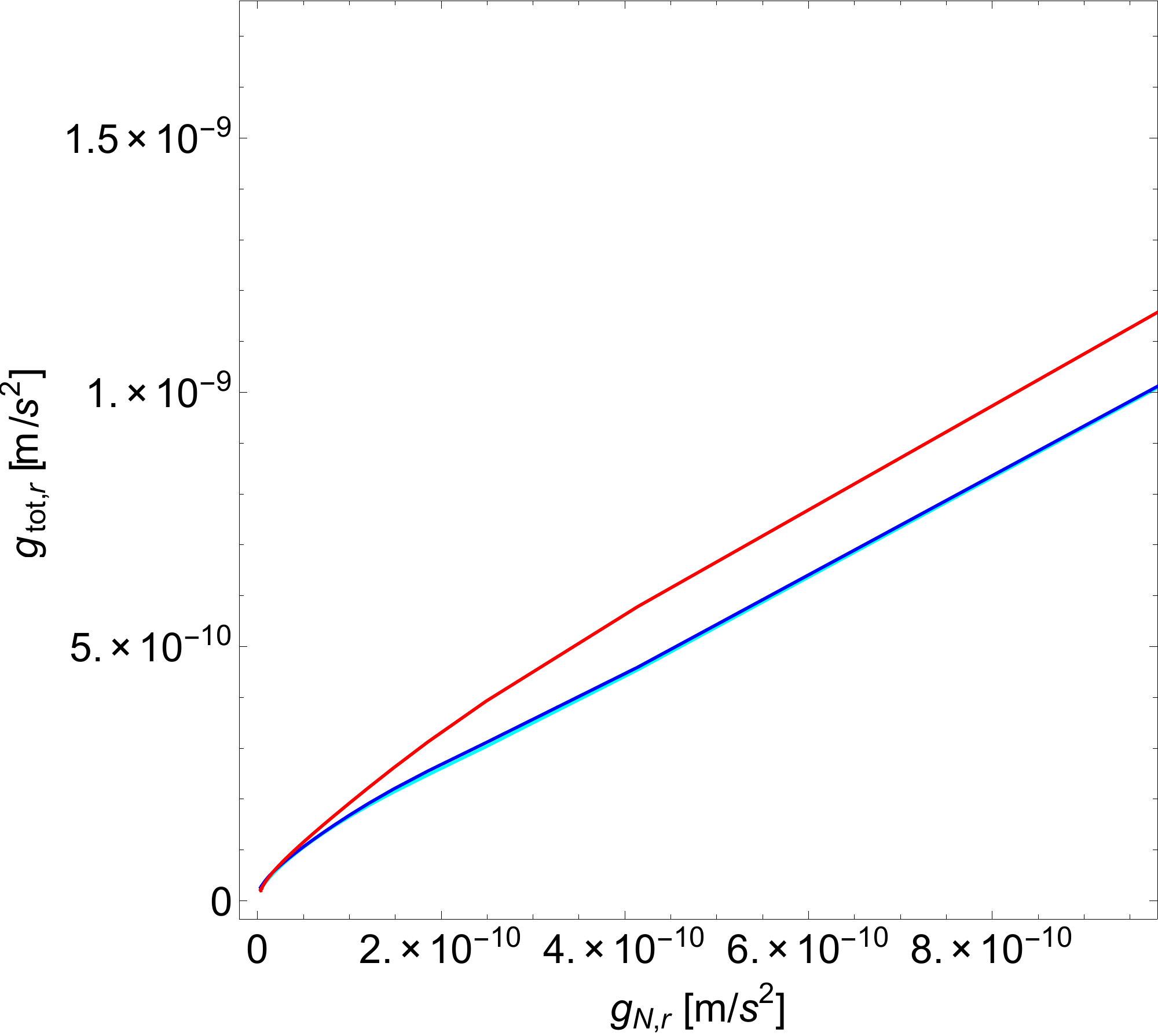}
	\includegraphics[width=0.65\columnwidth]{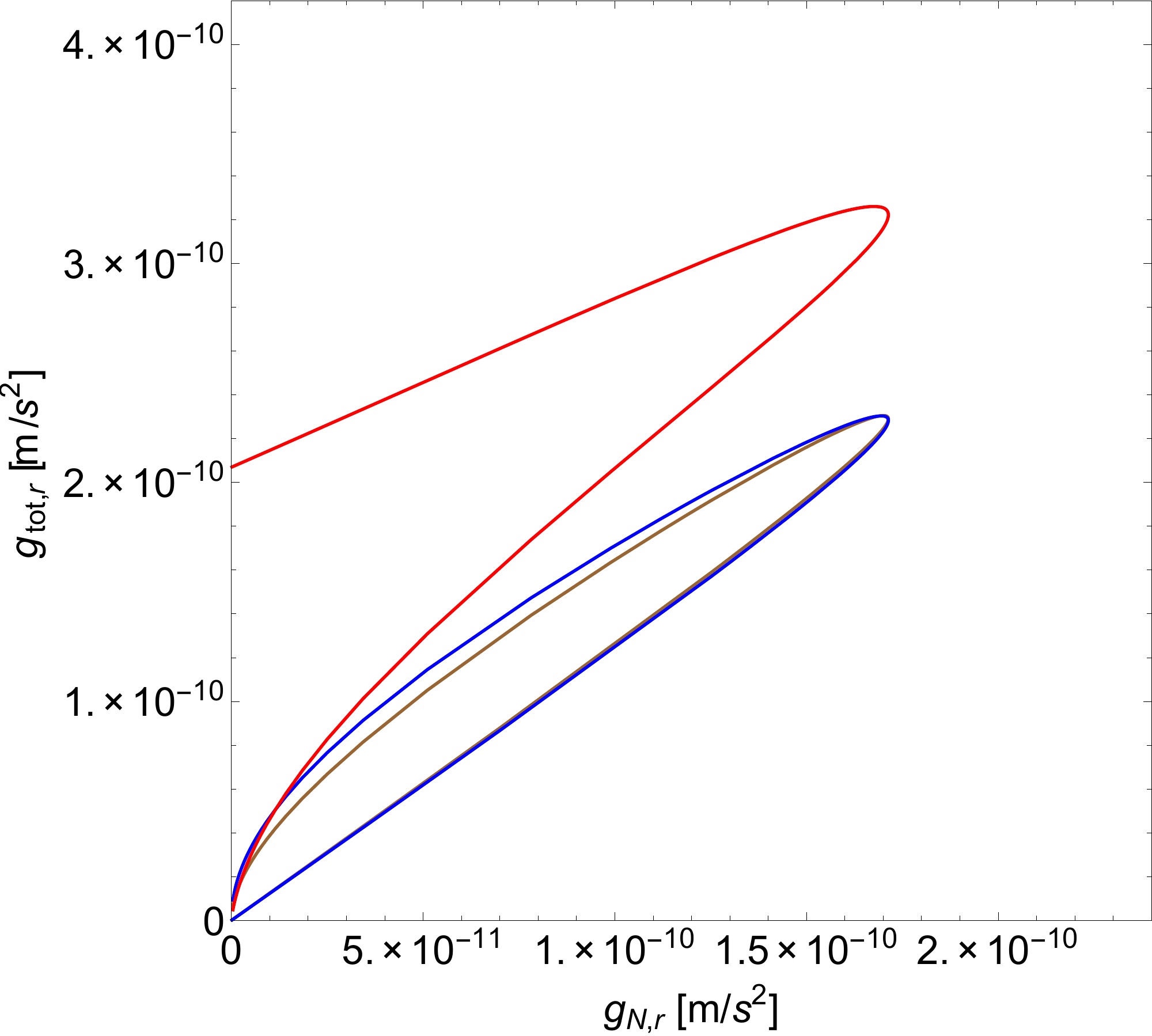}
	\includegraphics[width=0.65\columnwidth]{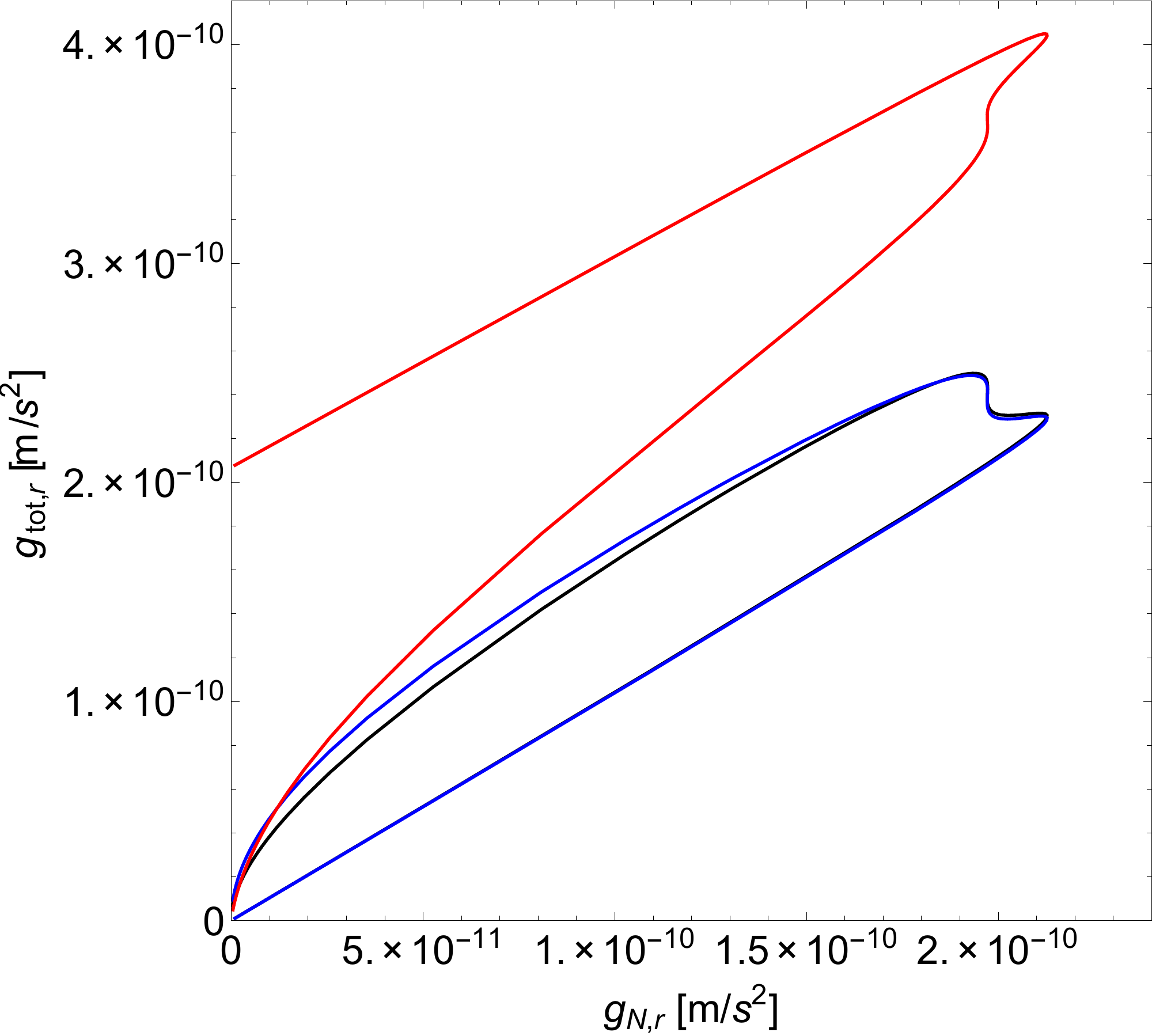}
	\caption{The $g2$-space geometry of Equation \eqref{potx}, and dark matter with the distributions of Equation \eqref{dm} (red is NFW and blue is isothermal), with the different baryonic distributions from Equations \eqref{18} and \eqref{19}. The difference between panels consist only in the baryonic matter; exponential disk (top panel), exponential disk and a de Vaucouleurs bulge (second panel from the top), Miyamoto-Nagai disk (third panel from the top) and Miyamoto-Nagai disk and Plummer bulge (bottom panel). The baryonic data is the same as used for Figure \ref{fig:1}. Additionally, $\rho_0=4\times 10^{-21}\,\frac{\text{kg}}{\text{m}^3}$, $\rho_1=2\times10^{-21}\,\frac{\text{kg}}{\text{m}^3}$, $R_0=3\,\text{kpc}$, $R_1=8\,\text{kpc}$ and the baryonic quantities are identical to those used for figure \ref{fig:1}.}
	\label{fig:3}
\end{figure}

Another point to make is in comparing the $g2$-space geometry of Equation \eqref{potx} to that of common dark matter models. Here, the pseudo-isothermal and NFW density profiles for dark matter will be considered.
\begin{equation}
	\begin{split}
		&\rho_{iso}(r,z)=\frac{\rho_0}{1+(\frac{\sqrt{r^2+z^2}}{R_0})^2},\\
		& \rho_{NFW}(r,z)=\frac{\rho_1}{\frac{\sqrt{r^2+z^2}}{R_1}(1+\frac{\sqrt{r^2+z^2}}{R_1})^2}.
	\end{split}
	\label{dm}
\end{equation}
Figure \ref{fig:3} illustrates the $g2$-space geometry of Equation \eqref{potx} and the dark matter distributions of Equation \eqref{dm}. It is clear that the primary differences between the models are present at small radii -- as discussed in \citet{Petersen:2017klw,Frandsen:2018ftj,Petersen:2019obe,Tian2019}. 
In the absence of a bulge, the solution to the missing gravity problem dictates the geometry at small radii. Finer details of the solution are clearly visible, as exemplified by the difference in geometry between the NFW and pseudo-isothermal dark matter geometries (without bugles). Given how sensitive the $g2$-space geometry is to the finer details of the solution to the missing gravity problem, it is remarkable how similar the geometry of Equation \eqref{potx} is to that of pseudo-isothermal dark matter.

\section{Summary and Discussion}
In this study, it has been shown how a toy model solution to the missing gravity problem (Equation \eqref{G}) can be reverse engineered from galactic rotation curve data. It has also been shown that this solution can be written in terms of the Newtonian centripetal acceleration from the baryonic matter only (Equation \eqref{potx}). The details of the toy model have been discussed in the context of its $g2$-space geometry and the results of \citet{Frandsen:2018ftj,Petersen:2017klw,Petersen:2019obe} obtained from the SPARC database \citep{Lelli:2016zqa}. The $g2$-space geometry highlights the finer details of the proposed solution to the missing gravity problem and as such provides an appropriate platform to discuss the details of a solution based on rotation curve data. The toy model is found to accommodate the most common $g2$-space geometries of the SPARC database, but not geometries which are consistent with some form of cuspy dark matter (nowhere geometries sampled at $r<r_{N}$ or leftward geometries). However, as recently proposed by \citet{Petersen:2019obe}; such geometries could be an artefact from a radial dependency of the mass to light ratios not accounted for in \citep{Frandsen:2018ftj}. In \citet{Petersen:2019obe} it is found that gas dominated galaxies follow a pronounced rightward geometry, consistent with some form of cored dark matter. Accepting this possibility, the toy model is consistent with all geometries present in the SPARC database.\newline
Lastly, the $g2$-space geometry of the toy model has been compared to those of NFW and pseudo-isothermal dark matter. This comparison shows a striking similarity between the toy model and isothermal dark matter to such a degree that one might consider the two loosely degenerate in terms of $g2$-space geometries. Extending this line of thought and collecting the results, the toy model shows that pseudo-isothermal dark matter can loosely be written in terms of Newtonian quantities derived from the baryonic matter (as far as the $g2$-space geometry is considered). This curious connection is an important take away from this study and should be viewed as a constraint for model builders. The toy model also suggest a low-acceleration departure form the predictions of general relativity, as also predicted by e.g. MOND modified gravity~\citep{Brada:1994pk}. However, whereas MOND modified gravity relies on an interpolation function, the toy model does not. A curious note is that the functional form of the toy model (Equation \eqref{potx}) is reminiscent of what is obtained from the post-Newtonian correction to general relativity, indicating that a mechanism to increase the magnitude of the post-Newtonian correction from general relativity for low accelerations may be worth considering.\bigskip

{\bf Acknowledgments:}
We would like to thank Indranil Banik for detailed feedback on draft. We also acknowledge partial funding from The Council For Independent Research, grant number DFF 6108-00623. The CP3-Origins center is partially funded by the Danish National Research Foundation, grant number DNRF90.\newline\newline

\bibliography{refs}

\end{document}